# Training recurrent neural networks with sparse, delayed rewards for flexible decision tasks

**Thomas Miconi**
The Neurosciences Institute
800 Silverado St, Suite 302
La Jolla CA 92037, USA
miconi@nsi.edu

**Abstract**

Recurrent neural networks in the chaotic regime exhibit complex dynamics reminiscent of high-level cortical activity during behavioral tasks. However, existing training methods for such networks are either biologically implausible, or require a real-time continuous error signal to guide the learning process. This is in contrast with most behavioral tasks, which only provide time-sparse, delayed rewards. Here we show that a biologically plausible reward-modulated Hebbian learning algorithm, previously used in feedforward models of birdsong learning, can train recurrent networks based solely on delayed, phasic reward signals at the end of each trial. The method requires no dedicated feedback or readout networks: the whole network connectivity is subject to learning, and the network's output is read from one arbitrarily chosen network cell. We use this method to successfully train a network on a delayed nonmatch-to-sample task (which requires memory, flexible associations, and non-linear mixed selectivities). Using decoding techniques, we show that the resulting networks exhibit dynamic coding of task-relevant information, with neural encodings of various task features fluctuating widely over the course of a trial. Furthermore, network activity moves from a stimulus-specific representation to a response-specific representation during response time, in accordance with neural recordings in behaving animals for similar tasks. We conclude that recurrent neural networks, trained with reward-modulated Hebbian learning, offer a plausible model of cortical dynamics during learning and performance of flexible association.

## Introduction

Recent evidence suggests that neural representations are highly dynamic, encoding multiple aspects of tasks, stimuli and commands in the joint fluctuating activity of entire populations of neurons, rather than in the stable activation of specific neurons (Meyers et al. 2008; Barak, Tsodyks, and Romo 2010; Stokes et al. 2013; Churchland et al. 2012). Models based on recurrent neural networks, operating in or near the chaotic regime, seem well-suited to capture similar dynamics (Jaeger 2001; Maass, Natschläger, and Markram 2002; Buonomano and Maass 2009; Sussillo and Abbott 2009). As a result, such models have been used to investigate the mechanisms by which neural populations solve various computational problems, including working memory (Barak et al. 2013), motor control (Sussillo et al. 2015), and selective evidence accumulation (Mante et al. 2013). However, the methods used to train these recurrent models are often not biologically plausible. The most common training methods require a teaching signal to be reproduced by the network, and adapt some or all of the network's connection weights by descending a gradient on the error between network output and teaching signal (Pearlmutter 1995; Jaeger 2001; Sussillo and Abbott 2009).

A more biologically plausible is based on so-called reward-modulated Hebbian learning: Hebbian synaptic modifications are modulated according to the value of a reward signal, making the cell more (or less) likely to fire for a given condition if its firing reliably predicts increased (or decreased) reward. This method, inspired by the effect of dopamine on synaptic plasticity, was successfully demonstrated and analyzed in *feedforward* spiking networks, often as a modification of spike-timing dependent plasticity (Izhikevich 2007; Florian 2007; Frémaux, Sprekeler, and Gerstner 2010). Importantly, these methods work even when rewards are time-sparse and delayed: they can solve the so-called "credit-assignment" problem, which arises when a single delayed reward signal must be apportioned to a sequence of actions taken before this reward (Izhikevich 2007; Frémaux, Sprekeler, and Gerstner 2010). However, unaided feedforward networks cannot generate the complex temporal dynamics observed in cortical activity and recurrent networks. Recently researchers have proposed to apply reward-modulated Hebbian learning to recurrent networks (Legenstein et al. 2010; Hoerzer, Legenstein, and Maass 2014). But these methods require a continuous, real-time reward-signal, based on instantaneous performance. This is in contrast with most behavioral tasks, both in nature and in the laboratory, where rewards are sparse, delayed, and not immediately ascribable to any specific previous action. To our knowledge there is no example of a biologically plausible learning algorithm for recurrent neural networks, that can successfully learn a given task using solely delayed, sparse reward signals.

Here we show that reward-modulated Hebbian learning can be used to train chaotic recurrent networks for flexible behaviors, with reward occurring in a delayed, one-time fashion after each trial, as in most animal training paradigms. This method is essentially identical to an algorithm previously described in a feedforward model of birdsong learning (Fiete and Seung 2006; Fiete, Fee, and Seung 2007) (see Methods). We apply this method to a simple flexible decision-making task, namely the sequential-XOR, which requires flexible stimulus-response associations, memory maintenance, and nonlinear selectivities. We show that the model reliably learns to perform this task. Investigating the way task-relevant aspects are represented in the network over time, we find that trained networks exhibit highly dynamic coding, as observed in recordings of animal frontal cortices. Furthermore, the system shows a shift from stimulus-dominated to decision-dominated encodings, flexibly "routing" stimulus representations to the adequate decision state, in accordance with observations. We conclude that reward-modulated Hebbian learning offers a plausible model of cortical learning, capable of building networks that dynamically represent and analyze stimuli and produce flexible responses in a way that is compatible with observed evidence.

**Methods**

*Model description*

Our model is a fully-connected continuous-time recurrent neural network of N neurons, governed by the following equations (Sompolinsky, Crisanti, and Sommers 1988; Sussillo and Abbott 2009; Jaeger 2001; Maass, Natschläger, and Markram 2002):

$$\tau \frac{dx_i}{dt} = -x_i(t) + \sum_{j=1}^{N} J_{j,i} r_j(t) + \sum_{k=1}^{I} B_{k,i} u_k(t) \quad \text{[Equation 1]}$$

$$r_i(t) = tanh(x_i(t)) \quad \text{[Equation 2]}$$

where $x_i$ is the activation of neuron i, $r_i$ is its actual response, $J_{j,i}$ is the connection weight from neuron j to neuron i, $u_k$ are the I inputs to the network, and $B_{k,i}$ is the connection weight from input k to neuron i. In addition, four arbitrarily chosen neurons have a constant activation x=1 and thus provide a bias input to other neurons. There is no separate feedback or output network. Instead, one cell in the network is arbitrarily designated as the "output" cell, and its response $r_0(t)$ at any given time is used as the network's response (this cell is otherwise entirely similar to all others). J is initialized with weights taken from a normal distribution with mean 0 and variance $g^2/N$, while the input weights $B_k$ are fixed and taken from a uniform distribution over the [-0.5,0.5] interval. For all simulations reported here, N=200, τ=10ms, and g=1.5; note that the latter value places the networks in the near-chaotic regime, where the long-term behavior will often remain non-periodic (Sompolinsky, Crisanti, and Sommers 1988).

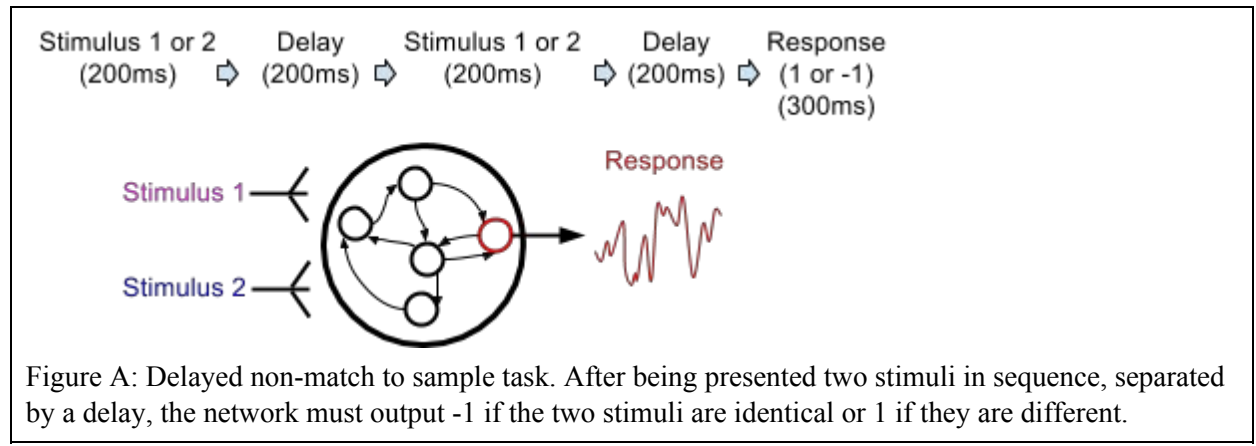

Figure A: Delayed non-match to sample task. After being presented two stimuli in sequence, separated by a delay, the network must output -1 if the two stimuli are identical or 1 if they are different.

Synapses between neurons are modified according to reward-modulated Hebbian learning. At the end of each trial period, a single-valued reward signal retroactively modulates potential weight changes accrued through Hebbian learning in the recent past, determining the sign and amplitude of the actual weight change. This principle, inspired by the effects of dopamine on synaptic plasticity, was successfully demonstrated in feedforward spiking networks (Izhikevich 2007; Frémaux, Sprekeler, and Gerstner 2010). In implementing reward-modulated Hebbian learning, we sought to remain close to the proven algorithm used in spiking networks, which led us to use time-sparse events. At any time, with a certain probability $F_p$ (which emulates the expected average spiking frequency of the neurons), the activation $x_i$ of each neuron i is subject to a small random fluctuation $\Delta x_i$ (which emulates small fluctuations in the timing of individual spikes). At this time, the neuron's synapses from other neurons are accrued a *potential* weight change (also called eligibility trace (Izhikevich 2007)) according to the following equation:

$$\Delta e_{j,i} = \Delta e_{j,i} + r_{j,i}(t) \Delta x_i(t)$$

Note that this is in accordance with standard Hebbian (covariance) learning, since the $r_{j,i}$ are the inputs to neuron i while $\Delta x_i$ describes its (above-expectation) activity. Also note that the eligibility trace for any synapse is cumulative, with new Hebbian events being added to the eligibility trace over a trial.

At the end of each trial, a certain reward R is issued to the network. From this reward, the system computes a *reward prediction error* signal, as observed in physiological experiments, by subtracting the expected reward for this trial $\overline{R}$ (see below for computation of $\overline{R}$). This reward-prediction signal is used to modulate the eligibility trace into an actual weight change:

$$\Delta J_{j,i} = \alpha \Delta e_{j,i}(R - \overline{R})$$

In our simulations, $F_p = 10$Hz, $\Delta x_i$ is taken from a normal distribution with mean 0 and standard deviation 0.02, $\alpha=0.03$. The eligibility trace for all synapses is reset to 0 at the start of each trial.

To gain a better intuitive understanding of reward-modulated Hebbian learning, we may consider its overall, simplified equation: $\Delta J(t) = r(t) \Delta x(t) (R-\underline{R})$. This equation is often understood by noting that the two first terms of the right-hand side essentially describe simple Hebbian covariance learning, on which (R-$\underline{R}$) acts as a modulator. However, another way to interpret the same equation is to group the last two terms of the right-hand side, $\Delta x(t) (R-\underline{R})$. Intuitively, this product stochastically computes the covariance between a neuron's activation at a given time and the obtained reward. If this term is positive (resp. negative), we want the neuron to fire more (resp. less) at that time. This can be done by adding (or subtracting) a multiple of the current inputs to the neuron (i.e. r(t)) to the synaptic weight, which makes the weight vector more (less) correlated with inputs at that time and thus increases (decreases) firing on the next occurrence of this particular input; this is precisely what this equation does.

To compute the reward prediction error signal (R-R), we need to estimate the expected reward R. Our method is essentially the same method as (Frémaux et al., 2010), which simply consists in maintaining a running average of recent rewards for a particular trial type (where trial type is determined by the combination of stimuli and expected response). As (Frémaux et al., 2010) pointed out, it is important that separate traces should be maintained for each trial type, so as to provide an accurate estimation of the expected reward R for each trial. Thus, after the n-th trial of a given type, R is updated as follows:

R(n) = trace R(n-1) + (1-trace) R(n)     [Equation 5]

Where R(n) is the reward for this trial, and R(n-1) was the expected reward after the previous trial of the same type. In all simulations, we set η (from Equation 4) to 0.03 and αtrace = 0.8.

*Task description: sequential-XOR*

The task considered here is the sequential-XOR - essentially a simple delayed nonmatch-to-sample task with only two stimuli. In every trial, we present two brief successive inputs to the network, with an intervening delay. These inputs can take either of two values A or B. The network's task is to determine whether the two inputs are identical (AA or BB, in which case the network should output -1) or different (AB or BA, in which case the network should output 1). We specify the input stimuli by using two different input channels $u_1$ and $u_2$; the identity of the input stimulus is determined by which channel is activated (i.e. for stimulus A, $u_1$=1 and $u_2$=0; for stimulus B, $u_1$=0 and $u_2$=1; remember that each input channel has an independent set of weights $B_{k,i}$ on all network neurons, as specified in Equation 1). In every trial, the first stimulus is presented for 200 ms, then after a 200 ms delay the second stimulus is presented for 200 ms. The trial goes on for an additional 500ms, thus each trial is 1100ms long. The network's response is determined by the activity of the output cell over the last 300 ms of the trial. The overall error E for this trial is the mean *absolute* difference between the network's output and the target response, over these last 300ms; the reward for the trial is simply minus the overall error, R=-E.

As explained above, we use two input channels to specify either of the two possible stimuli. However, we also explored different encodings, such as using two different values over a single input channel (A: $u_1$=-1; B: $u_1$=1) or using one channel to transmit the first stimulus and another channel to transmit the second stimulus. All produced successful results. In this paper we show results for the separate-channels representation, since this seems most relevant to the situation of frontal cortices in which different stimuli would be represented by inputs from different population of neurons.

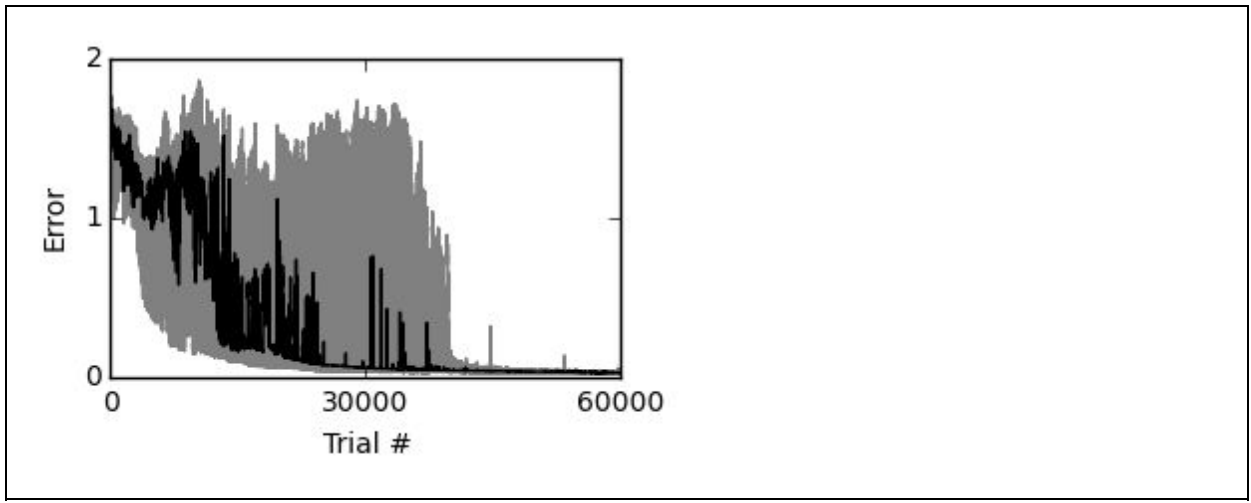

Figure 1. Time course of the error as a function of the number of trials (dark line: median over 10 trials; gray area: inter-quartile range). The quantity reported here is the current maximum error, at any time, across all four possible stimulus combinations. The network reliably learns to perform the task within about 30000 trials.

**Results**

We trained our networks to perform a simple flexible decision task, namely, the sequential-XOR task. The network is exposed to two 200ms stimuli, separated by a 200ms delay. These stimuli can be of either of two types (that is, activate one of two input channels to the network), which we call A and B. The network's task is to indicate whether the two successive stimuli were identical (A then A or B then B), in which case the network's output cell must respond with -1; or different (A then B or B then A), in which case the output cell must respond with 1. The response is evaluated over the last 300ms of the 1100ms trial (the "response" period). At the end of each trial, the network receive a single reward signal, based on the total error for this trial (i.e. the mean absolute difference between output cell response and target response over the response period). This reward signal is then used to modulate the Hebbian synaptic traces accumulated over the trial into actual synaptic weight changes.

This simple task exhibits several interesting features. First, it is arguably the simplest possible flexible decision task: on seeing the second stimulus, the network must flexibly produce a different response based on the identity of the first stimulus. Second, because the intervening delay is much longer than the neural time constant, the network must implement some memory of the first stimulus before the second stimulus arises. Third, to solve this task, some cells in the network (including the output cell) must necessarily possess some form of nonlinear mixed selectivity, a hallmark of neural responses in prefrontal cortices (Rigotti et al. 2013).

The networks consistently learn to perform the task with high accuracy. Figure 1 shows the time course of the median max-error (largest error across all four possible stimulus combinations) for 10 training runs, each starting with a different randomly initialized network. Shaded area indicates 1st and 3rd quartile over the 10 runs. All runs converge to very low residual error within 50000 trials. The frequent instabilities in the early phase of learning reflect the near-chaotic regime of the networks, in which small weight changes will occasionally produce large changes in response; nevertheless, the reliable convergence to negligible error demonstrates that reward-based learning can consistently overcome this instability.

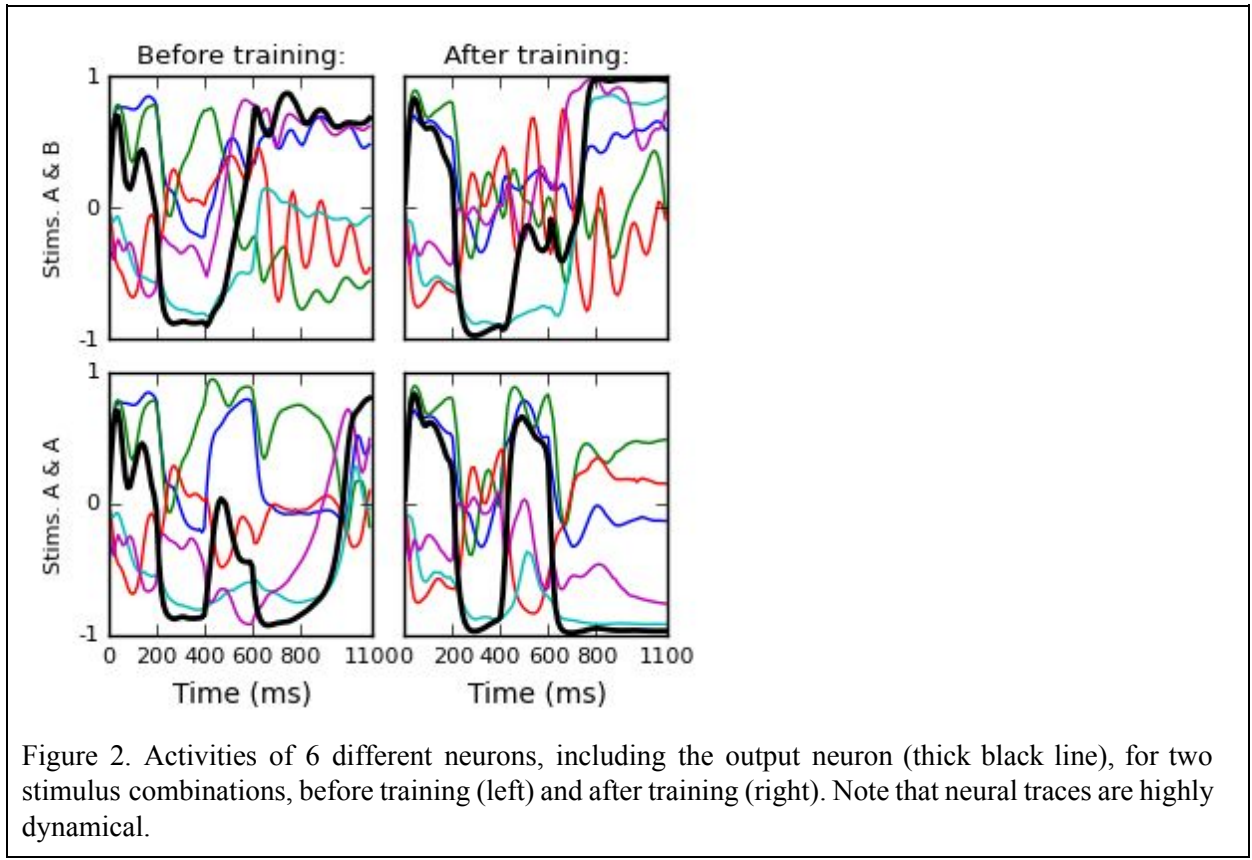

Figure 2. Activities of 6 different neurons, including the output neuron (thick black line), for two stimulus combinations, before training (left) and after training (right). Note that neural traces are highly dynamical.

How does the network represent and maintain traces of incoming stimuli? One possibility is that certain neurons encode stimulus identity by maintaining a stable "register" value over time, such that a certain cell response directly specifies stimulus identity in a relatively time-independent manner. By contrast, physiological studies suggest that neural coding during working memory task is highly dynamical, with stimulus identity being represented by widely fluctuating cell activations, in such a way that the tuning of individual neurons significantly changes over the course of a trial (Meyers et al. 2008; Barak, Tsodyks, and Romo 2010; Stokes et al. 2013). To investigate the encoding and maintenance of stimulus identity over time in the network, we used the temporal cross-classification approach (Meyers et al. 2008; Stokes et al. 2013). We trained a classifier to decode various task-relevant features (identity of 1st and 2nd stimulus, final response) based on population activity at a given time. When the population data used for training and decoding are sampled at the same time point in the trial (always using separate data for training and testing), this method can detect whether the network encodes information about a specific feature at that time. In addition, by collecting training data and decoding data at different time points, we can determine whether the representation used by the network is stable across these two time points - that is, whether the activity patterns used to represent a certain feature are the same at these two time points - or whether, on the contrary, the network uses different encoding representations from one time point to the next.

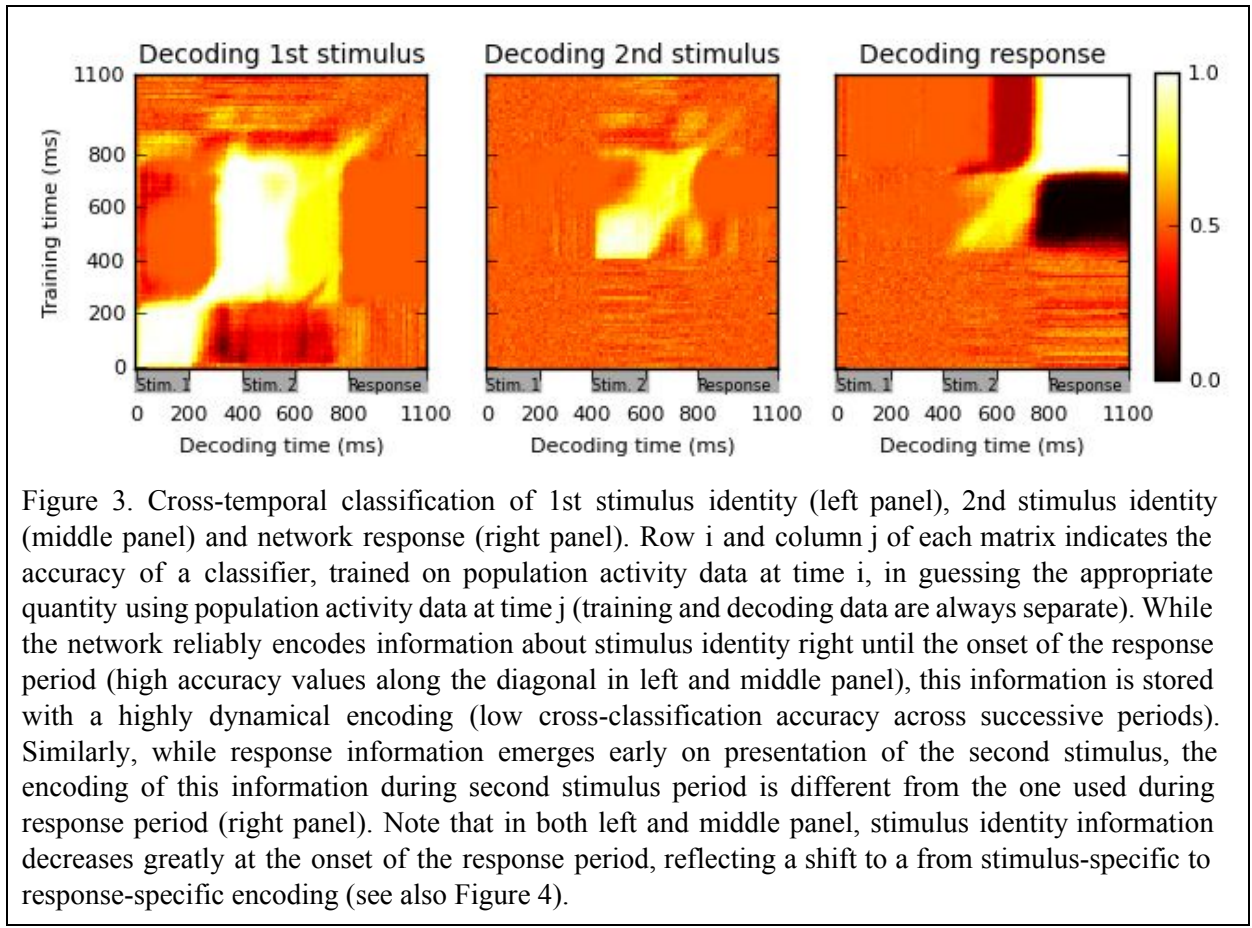

Figure 3. Cross-temporal classification of 1st stimulus identity (left panel), 2nd stimulus identity (middle panel) and network response (right panel). Row i and column j of each matrix indicates the accuracy of a classifier, trained on population activity data at time i, in guessing the appropriate quantity using population activity data at time j (training and decoding data are always separate). While the network reliably encodes information about stimulus identity right until the onset of the response period (high accuracy values along the diagonal in left and middle panel), this information is stored with a highly dynamical encoding (low cross-classification accuracy across successive periods). Similarly, while response information emerges early on presentation of the second stimulus, the encoding of this information during second stimulus period is different from the one used during response period (right panel). Note that in both left and middle panel, stimulus identity information decreases greatly at the onset of the response period, reflecting a shift to a from stimulus-specific to response-specific encoding (see also Figure 4).

Results in Figure 3 suggest highly dynamical representation of stimuli by the network. For example, the identity of the first stimulus can be successfully decoded during both first and second stimulus presentation, as well as during the intervening delay (as shown by high accuracy values on the diagonal during this entire period). Within either of the two stimulus presentation periods, coding is relatively stable, as indicated by the square-like shapes of high-accuracy cross-classification "islands" within each presentation period. However, the cross-classification between these two periods (as seen in the areas of the matrix corresponding to one stimulus presentation in the training time, and the other stimulus presentation period in the decoding time) is essentially at chance level. This suggests that while the network maintains information about 1st-stimulus identity across the first 800ms of the trial, the way in which this identity is represented changes widely between successive periods within the trial. Similarly, 2nd-stimulus identity is maintained from 2nd stimulus onset until the beginning of the response period, but in a dynamical manner (low cross-classification accuracy between the 400-600ms period and the 600-800ms period).

Another feature of these plots is that the accuracy of stimulus identity decoding is strongly reduced during the "response" period (the 800-1100ms period, over which the output cell is evaluated for the error signal). In this period, even along the diagonal, decoding accuracy for the identity of either first or second stimulus diminishes greatly in comparison to previous periods. This suggests that the network largely stops maintaining information about the specific identity of previous stimuli, and instead encodes solely the actual response, as shown by the very strong classification accuracy in the third plot. Network response is also encoded in a dynamic period over time - indeed, network activity during the second stimulus period seems to anti-predict the actual response (dark patches in the 3rd plot of Figure 3), suggesting that response encoding during the 2nd stimulus period is somehow anti-correlated with its encoding during the response period. However, just before the onset of the response period, the networks implements a highly stable representation of the response, as seen from the almost perfect rectangular zone of high cross-classification accuracy in the top-right corner of the 3rd plot. Importantly, this does not imply that all (or even most) neurons enter a stable, 'frozen' activity state (Figure 2 shows that individual neural activities remain dynamic during the response period). Rather, it means that enough neurons have sufficiently stable responses over that time that network response can be accurately decoded by using the same comparisons over the entire response period.

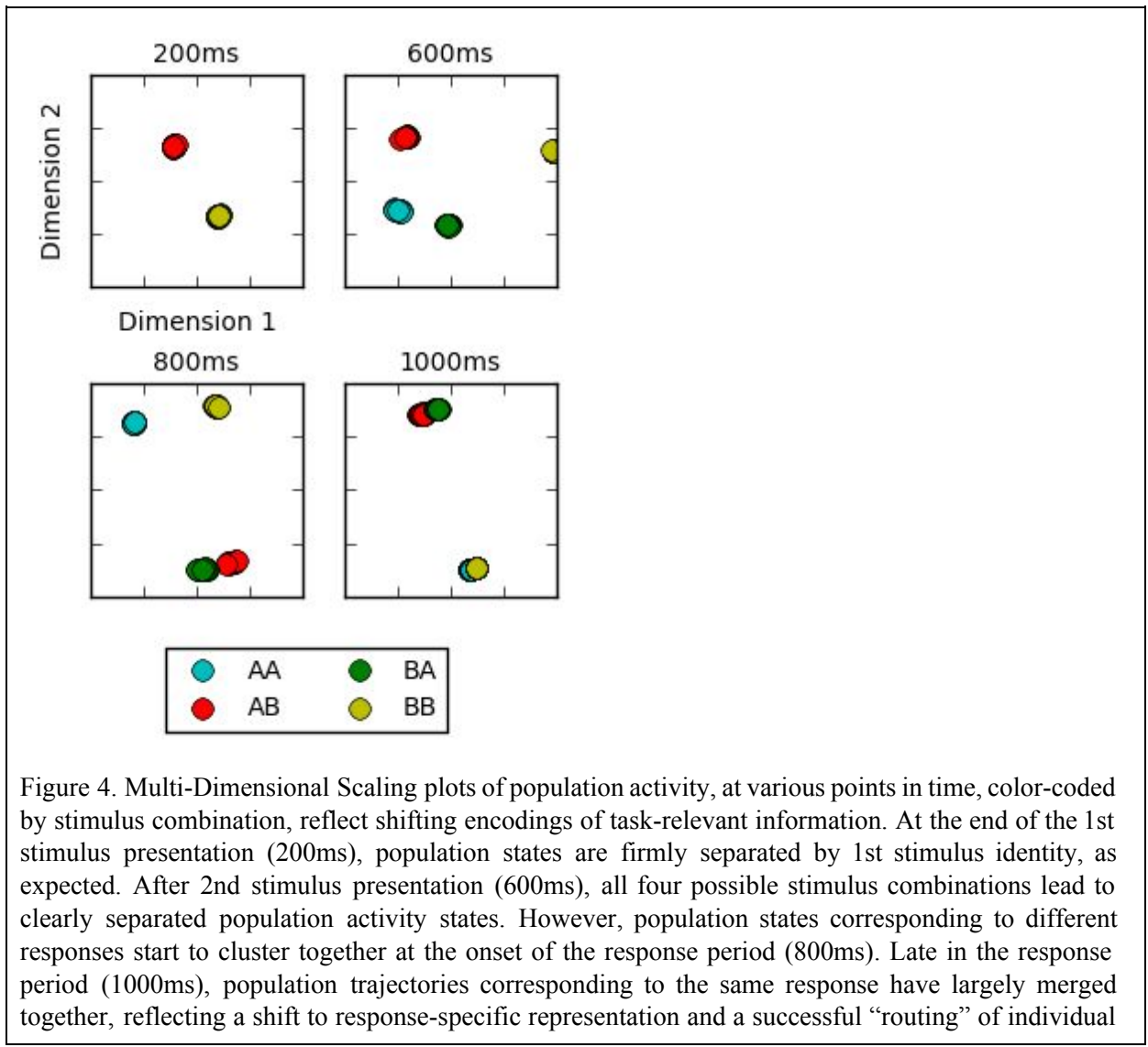

Figure 4. Multi-Dimensional Scaling plots of population activity, at various points in time, color-coded by stimulus combination, reflect shifting encodings of task-relevant information. At the end of the 1st stimulus presentation (200ms), population states are firmly separated by 1st stimulus identity, as expected. After 2nd stimulus presentation (600ms), all four possible stimulus combinations lead to clearly separated population activity states. However, population states corresponding to different responses start to cluster together at the onset of the response period (800ms). Late in the response period (1000ms), population trajectories corresponding to the same response have largely merged together, reflecting a shift to response-specific representation and a successful "routing" of individual

> stimulus-specific states to the adequate response-specific state.

The fact that the network mostly "forgets" specific stimulus identity during the response period suggests that the population moves from a so-called "stimulus-specific" encoding to a "response-specific" encoding: the stimulus-specific response is flexibly routed to the appropriate, context-dependent response state, as previously observed in cortical activity during a flexible association task (Stokes et al. 2013). To test this interpretation, following (Stokes et al. 2013), we produce Multi-Dimensional Scaling (MDS) plots of population activity, at different time points, for different stimulus conditions (Figure 4). MDS attempts to find the 2D projection such that the distance between any two data points are as similar as possible to their actual distances in the full-dimensional space: nearby (distant) population states should produce nearby (distant) points on the MDS plot.

During first stimulus presentation (200ms), population trajectories are grouped according to 1st stimulus identity, as expected, since this is the only information available to the network at that time. After second stimulus presentation, population trajectories have split again, in such a way that all possible stimulus identity combinations generate different, consistent trajectories. This four-way distinction begins to erode at the onset of the response period (800ms), in which the population states for "different-identity" stimulus combinations (A-B and B-A) begin to overlap. By the late response period (1000ms), the trajectories corresponding to either response ("same" or "different") have essentially merged together, largely erasing any distinction based on specific identity of either first or second stimulus. Thus, during the response period, the network flexibly moves from a stimulus-specific representation to a response-specific representation, consistent with physiological observations.

**Discussion**

In this paper we introduced a novel, biologically plausible learning algorithm that can fashion a recurrent neural network to learn a flexible association task, using only time-sparse, delayed rewards to guide learning. Our algorithm is largely an adaptation (and perhaps a unification) of previous efforts in spiking feedforward networks, and continuous-time recurrent networks with real-time, continuous supervised feedback. The trained network exhibit the dynamic, population-wide encoding of task-relevant information observed in neural recordings (Meyers et al. 2008; Stokes et al. 2013; Barak, Tsodyks, and Romo 2010)

The flexible, dynamic coding observed in prefrontal activity has led to suggestions that cortex implements 'silent' memory traces by using short-term synaptic plasticity (Barak, Tsodyks, and Romo 2010; Stokes 2015). Short-term synaptic plasticity clearly plays an important role in neural responses, and may well play an important role in maintaining a "hidden internal state" of the network (Buonomano and Maass 2009). However, our network does not implement short-term synaptic plasticity; no weight modification occurs during trials, and all the decoding results reported above were obtained with frozen synaptic weights. Our results suggest that the highly dynamic activities spontaneously produced by near-chaotic recurrent networks can be harnessed to produce the dynamic encodings observed in experiments, using only sparse, delayed rewards and biologically plausible plasticity rules. Thus, while short-term synaptic plasticity clearly affects neural responses, it may not be required to explain the highly dynamic nature of working-memory encodings.

It is unlikely that cortical connectivity should be drastically and finely remodeled through a long training process for any new task. For example, while monkeys require extensive training to perform decision tasks, human subjects can quickly perform new tasks simply by verbal instruction. Rather, it is more likely that the long process of slow reward-modulated synaptic modification describes the process of cortical molding in early development, whereby specific cortical areas slowly learn to perform a generic *type* of task, as determined by which other areas they receive information from, and by their specific pattern of reward-based plasticity modulation (for example, certain cortical areas, but not others, receive specific dopamine input on aversive trials, which is critical for aversive learning (Lammel et al. 2012)). Then, to perform a specific task, it would be sufficient to learn to provide the proper inputs to these generic networks. The latter process of flexible task specification through input modulation is likely to involve not just other cortical areas, but also the basal ganglia and (tonic) dopamine delivery. While these external components were unnecessary for the simple, single-task settings described in this paper, elucidating the interactions between cortical, limbic and dopaminergic structures is an important future task for the study of flexible behavior and its neural implementation.